# Surface tide on a rapidly rotating body

## Xing Wei⋆

*Institute of Natural Sciences, School of Mathematical Sciences, Shanghai Jiao Tong University, Shanghai 200240, China*

### ABSTRACT

By solving Laplace's tidal equations with friction terms, we study the surface tide on a rapidly rotating body. When $\epsilon = \Omega^2 R/g$ (i.e. the square of the ratio of the dynamical time-scale to the rotational time-scale) is very small for the Earth, the asymptotic result is derived. When it is not so small (e.g. for a rapidly rotating star or planet), we perform numerical calculations. It is found that when rotation is sufficiently fast ($\epsilon$ reaches 0.1), a great amount of tidal resonances appear. To generate the same level of tide, a faster rotation corresponds to a lower tidal frequency. Friction suppresses tidal resonances but it cannot completely suppress them at fast rotation. The thickness of the fluid layer can change tidal resonances but this change becomes weaker at faster rotation. This result can help us to understand tides in the atmosphere of a rapidly rotating star or planet, or in the ocean of a neutron star.

**Key words:** hydrodynamics – waves – binaries: general.

## 1 INTRODUCTION

The study of the tidal problem has a long history. Newton formulated the tidal force, which is the gravitational perturbation exerted by companion on primary that is inversely proportional to the cube of distance. A century later, Laplace established a dynamic theory. He considered a thin ocean layer covering the Earth. In his theory, the Earth's rotation and the motion of the free surface are considered, so that the tide behaves like a wave motion, called a tidal wave (Souchay, Mathis & Tokieda 2013). Laplace's theory is mathematically described by Laplace's tidal equations (i.e. a two-dimensional shallow-water model in which the ocean is thin compared to the Earth's radius; Pedlosky 1987). In Laplace's theory, gravity and rotation are the two major factors. Consequently, a tidal wave is a mixed mode of a surface gravity wave (e.g. ripples on a lake surface; Lighthill 1978) and an inertial wave (e.g. a fluid column on top of an obstacle in a rotating tank; Greenspan 1968). In astronomy, the former is called the *f* mode and the latter is called the *r* mode.

In this paper, we apply Laplace's theory to the study of the surface tide on rapidly rotating stars and planets. This model can be used to interpret the tide raised in stellar and planetary atmospheres or in the ocean on a neutron star. The surface tide of a neutron star might be not very important because of its small size (Piro & Bildsten 2005), but the eigenmodes in the ocean of a neutron star can be found by solving Laplace's tidal equations. These eigenmodes can be used to interpret the star's oscillation signals (Ho & Lai 1999). The surface tide is a dynamical tide that differs from an equilibrium tide of hydrostatic balance. The internal dynamical tide has been extensively studied by, for example, Cowling (1941), Zahn (1970, 1975, 1977), Goldreich & Nicholson (1989), Lai (1997), Ogilvie & Lin (2004), Wu (2005), Lai & Wu (2006), Goodman & Lackner (2009), Fuller & Lai (2012), Ogilvie (2014), Fuller, Luan & Quataert (2016), Xu & Lai (2017), Wei (2016b, 2018) and Lin & Ogilvie (2018). The surface tide has also been studied (e.g. Zhugzhuda 1982; Ho & Lai 1999; Chirenti, de Souza & Kastaun 2015; MacLeod et al. 2018). However, the surface tide in the regime of rapid rotation has not been well studied.

With regards to Laplace's theory, the free-oscillation problem in the absence of a tidal force has been studied by Longuet-Higgins (1968) in two asymptotic limits (i.e. the regimes of very slow and very fast rotation). The forced oscillation problem was also qualitatively studied in the same paper, and the condition of tidal resonance was briefly derived but the solutions were not given. For the study of the surface tide, the parameter $\epsilon = \Omega^2 R/g$, which is the square of the ratio of rotational frequency $\Omega$ to dynamical frequency $\sqrt{g/R}$, is used to measure the relative strength of rotation and surface gravity. For the Earth, $\epsilon \approx 3.44 \times 10^{-3}$ is very small. However, some stars and planets rotate so rapidly that $\epsilon$ is large; for example, $\epsilon \sim 0.1$ for Jupiter and $\alpha$ Eridani, and $\epsilon$ ranges from 0.1 to 1 for neutron stars. When rotation is fast, the Coriolis force couples different modes in the latitude direction such that numerical calculations become inevitable. In this paper, we study this regime of rapid rotation. We establish the model in Section 2, we discuss the results in Section 3 and we give a brief summary in Section 4.

⋆ E-mail: xing.wei@sjtu.edu.cn

## 2 MODEL

We now apply Laplace's two-dimensional shallow-water model to the study of the surface tide of rapidly rotating stars and planets. First, we need to justify the validity of the model for stars and planets. One question is whether a model of incompressible fluid can be applied to the study of compressible fluid in stars and planets. We are concerned with the tidal wave. The natural frequency of a tidal wave is much lower than that of a sound wave in compressible fluid. Therefore, a sound wave cannot interact with a tidal wave and the incompressible fluid model can be applied to the study of tidal waves. Another question is then whether the spherical geometry of Laplace's model can be applied to a rapidly rotating body with a spheroidal figure. The diagram of $\epsilon \sim \Omega^2/(\pi G \bar{\rho})$ versus eccentricity can be found in Chandrasekhar (1969) and Lai, Rasio & Shapiro (1993). For example, for Jupiter, $\epsilon \approx 0.1$ and its flattening $\approx 0.06$, so the spherical geometry seems fine for $\epsilon \lesssim 0.1$. For a neutron star, $\epsilon$ can reach the order of unity but the spherical model can be applied because it is very rigid (here, 'rigid' refers to its equation of state). However, although it is not very rigorous for a very rapidly rotating body, the spherical model can provide us with some information about tidal resonances.

Suppose that a thin fluid layer with a uniform depth covers the surface of a rotating body. We apply Laplace's tidal equations in the frame rotating with the primary (Souchay et al. 2013) by adding the following friction terms:

$$\frac{\partial u_\theta}{\partial t} - 2\Omega \cos\theta\, u_\phi = \frac{1}{R}\frac{\partial}{\partial \theta}(-g\eta + V + \Psi) - \gamma u_\theta;$$
$$\frac{\partial u_\phi}{\partial t} + 2\Omega \cos\theta\, u_\theta = \frac{1}{R\sin\theta}\frac{\partial}{\partial \phi}(-g\eta + V + \Psi) - \gamma u_\phi;$$
$$\frac{\partial \eta}{\partial t} + \frac{h}{R\sin\theta}\left[\frac{\partial}{\partial \theta}(u_\theta \sin\theta) + \frac{\partial u_\phi}{\partial \phi}\right] = 0. \tag{1}$$

Here, the first two equations are momentum conservation and the third equation is mass conservation. These equations are in spherical polar coordinates $(r, \theta, \phi)$. $R$ is the radius, $h$ is the thickness of the fluid layer, $\Omega$ is the rotational frequency, $g$ is the surface gravity, $u_\theta$ and $u_\phi$ are the colatitude and longitude velocities, respectively, $\eta$ is the surface displacement, $V$ is the tidal potential and $\Psi$ is the perturbation of the self-gravity potential caused by $\eta$. The two terms proportional to $2\Omega$ are the Coriolis force due to rotation. We assume that the frictional force is proportional to the velocity and that the coefficient $\gamma$, which is the so-called Rayleigh drag that was adopted by Ogilvie (2009), is constant.

The major component of the tidal potential is the semi-diurnal potential

$$V = \frac{1}{4}gR\frac{m_2}{m_1}\left(\frac{R}{d}\right)^3(\cos^2\delta)\,P_2^2(\cos\theta)\cos(2\phi - 2\omega t), \tag{2}$$

where $m_1$ and $m_2$ are the mass of the primary and the companion, respectively, $d$ is the distance, $\delta$ is the companion's declination, $\omega = \omega_o - \Omega$ is the tidal frequency in the frame rotating with the primary, where $\omega_o$ is the orbital frequency, and $P_2^2$ is the associated Legendre polynomial.

According to the expression of tidal potential in equation (2), the variables $u_\theta$, $u_\phi$, $\eta$ and $\Psi$ can be expanded with spherical harmonics and the complex spectral coefficients are denoted by $\hat{u}_\theta$, $\hat{u}_\phi$, $\hat{\eta}$ and $\hat{\Psi}$, e.g.

$$\eta = \Re\left[\sum_{n,m}\hat{\eta}P_n^m(\cos\theta)\,e^{im(\phi - \omega t)}\right]. \tag{3}$$

Here, $n$ is the colatitude wavenumber, $m$ is the longitude wavenumber and $\Re$ denotes the real part of a complex variable. The perturbation of the self-gravity potential is related to the surface displacement through $\hat{\Psi} = 3(\rho/\bar{\rho})g\hat{\eta}/(2n+1)$, where $\rho$ is the fluid density of the thin layer and $\bar{\rho}$ is the mean density of the primary.

To find more general physics, we solve the dimensionless equations. Normalizing length with $R$, time with rotational time-scale $\Omega^{-1}$, velocity with $\Omega R$, tidal potential with $gR$ and $\gamma$ with dynamical frequency $\sqrt{g/R}$ (note that we vary $\Omega$ in the numerical calculations so it is not proper to normalize $\gamma$ with $\Omega$), we are led to the following dimensionless equations:

$$\epsilon\frac{\partial u_\theta}{\partial t} + \sqrt{\epsilon}\gamma u_\theta - 2\epsilon\cos\theta\, u_\phi + a\frac{\partial \eta}{\partial \theta} = \frac{\partial V}{\partial \theta};$$
$$\epsilon\frac{\partial u_\phi}{\partial t} + \sqrt{\epsilon}\gamma u_\phi + 2\epsilon\cos\theta\, u_\theta + \frac{a}{\sin\theta}\frac{\partial \eta}{\partial \phi} = \frac{1}{\sin\theta}\frac{\partial V}{\partial \phi};$$
$$\frac{\partial \eta}{\partial t} + \frac{h}{\sin\theta}\left[\frac{\partial}{\partial \theta}(u_\theta \sin\theta) + \frac{\partial u_\phi}{\partial \phi}\right] = 0. \tag{4}$$

The parameter $\epsilon = \Omega^2 R/g$ measures the relative strength of rotation, $\gamma$ measures friction and $a = 1 - 3(\rho/\bar{\rho})/(2n+1)$ arises from $\Psi$, which is of the order of unity. The dimensionless tidal frequency is $\omega = \omega_o/\Omega - 1$.

In equations (4), the longitude wavenumber $m$ can be decoupled but the Coriolis force couples the colatitude wavenumber $n$, such that we have to solve equations (4) numerically. We decouple $m$ and take the sum of $n$. Because $P_n^m$ for $n < m$ is 0, the sum starts from $m$ and

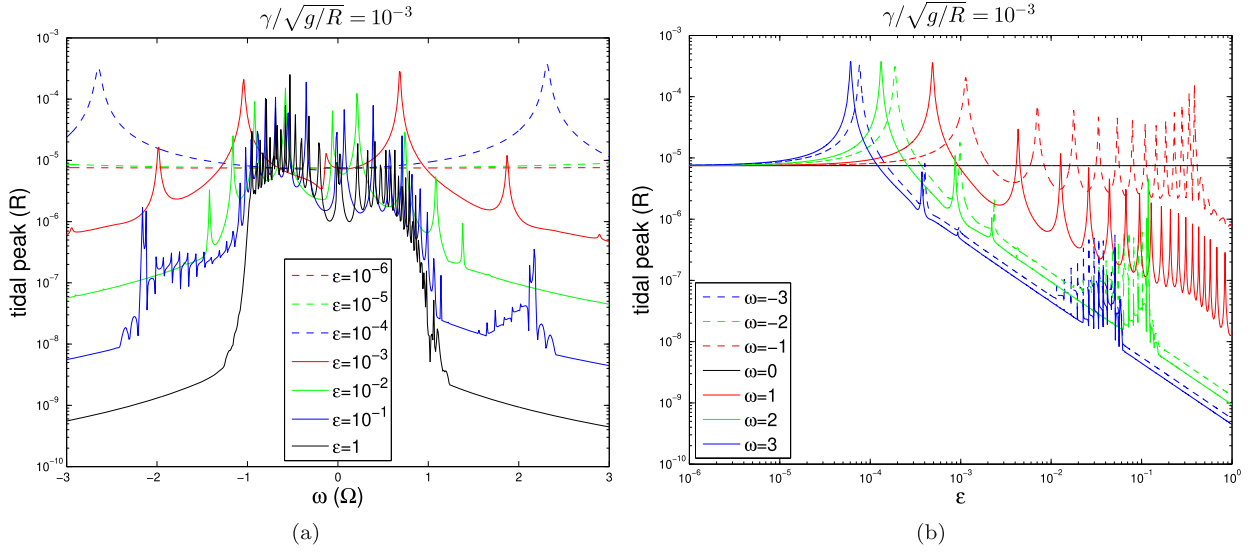

**Figure 1.** (a) Tidal peak versus $\omega$ with different values of $\epsilon$. (b) Tidal peak versus $\epsilon$ at $\gamma/\sqrt{g/R} = 10^{-3}$ with different $\omega$.

ends at $m + N - 1$. Thus, we obtain a linear system

$$\sum_{n=m}^{m+N-1} \left[ (\sqrt{\epsilon}\gamma - \mathrm{i}\epsilon m\omega)P_n^m \hat{u}_\theta - 2\epsilon\cos\theta\, P_n^m \hat{u}_\phi + a\frac{\mathrm{d}P_n^m}{\mathrm{d}\theta}\hat{\eta} \right] = \hat{V}\frac{\mathrm{d}P_{n_0}^m}{\mathrm{d}\theta}, \tag{5}$$

$$\sum_{n=m}^{m+N-1} \left[ 2\epsilon\cos\theta\, P_n^m \hat{u}_\theta + (\sqrt{\epsilon}\gamma - \mathrm{i}\epsilon m\omega)P_n^m \hat{u}_\phi + \frac{\mathrm{i}ma}{\sin\theta}P_n^m \hat{\eta} \right] = \frac{\mathrm{i}m}{\sin\theta}\hat{V}P_{n_0}^m,$$

$$\sum_{n=m}^{m+N-1} \left[ \frac{h}{\sin\theta}\frac{\mathrm{d}}{\mathrm{d}\theta}(\sin\theta\, P_n^m)\hat{u}_\theta + \frac{\mathrm{i}hm}{\sin\theta}P_n^m \hat{u}_\phi - \mathrm{i}m\omega P_n^m \hat{\eta} \right] = 0,$$

controlled by the following parameters: $\varepsilon$ of rotation, $\gamma$ of friction, $h$ of fluid layer, $\rho/\bar{\rho}$ of fluid density, and $\hat{V}$, $n_0$, $m$ and $\omega$ of the tidal potential. Next, we collocate equations (5) on the $N$ zeros of $P_n^m$ to obtain a linear system $Ax = b$, where $A$ is a $3N$-by-$3N$ matrix. To justify the convergence of the solution, we increase $N$ until the relative error of $|x|^2$ is less than 1 per cent.

As expressed in equation (3), the tide is a travelling wave in the longitude direction, and the tidal amplitude is a smooth function of colatitude $\theta$, that is, the superposition of $P_n^m(\cos\theta)$. We are concerned with the summit of the tidal amplitude, $\eta_{max}$, which is called the tidal peak. We output the tidal peak using our numerical calculations.

## 3 RESULTS

In equations (4), we take some typical values: $h = 10^{-3}$, $\rho/\bar{\rho} = 1$ for a homogeneous fluid sphere and the amplitude of the tidal potential $\hat{V} = 10^{-6}$. Declination $\delta$ in equation (2) is set to be 0 because we are concerned with the tidal peak. We take four values for $\gamma/\sqrt{g/R}$: $10^{-4}$, $10^{-3}$, $10^{-2}$ and $10^{-1}$. The results of $\gamma/\sqrt{g/R} < 10^{-4}$ are similar to the results of $\gamma/\sqrt{g/R} = 10^{-4}$ (i.e. the asymptotic results). However, it is difficult to estimate the realistic $\gamma$, which is related to turbulent viscosity, because the way in which turbulence is suppressed by the tide is controversial (Goldreich & Nicholson 1977; Zahn 1977; Goodman & Oh 1997). We calculate $\varepsilon$ up to 1. The frequency of the inertial wave is within twice the rotation frequency (i.e. from $-2\Omega$ to $2\Omega$), and we calculate the dimensionless tidal frequency $\omega$ from $-3$ to 3. We show only the results of the semi-diurnal tide and the other tidal components have similar results.

Fig. 1(a) shows the tidal peak versus $\omega$ at $\gamma/\sqrt{g/R} = 10^{-3}$ with different values of $\epsilon$. At slow rotation ($\epsilon = 10^{-6}$ and $10^{-5}$), there are no obvious tidal resonances. When the rotation becomes faster (i.e. $\epsilon$ increases from $10^{-4}$ to 1), more and more tidal resonances appear. At $\epsilon = 0.1$ and 1, we find a large amount of tidal resonances. At $\epsilon = 0.1$, the tidal resonances are almost within $\omega = -2$ and 2 while at $\epsilon = 1$ they concentrate within $\omega = -1$ and 1. We can conclude that faster rotation leads to more tidal resonances within a narrower range of tidal frequency.

Fig. 1(b) shows the tidal peak versus $\epsilon$ at $\gamma/\sqrt{g/R} = 10^{-3}$ with different $\omega$. Similar to Fig. 1(a), it shows that slow rotation cannot lead to tidal resonances whereas fast rotation can. It also provides some new information. First, tidal resonances still occur at $\omega = \pm 3$, which is already out of the frequency range of the inertial wave. This is not surprising because, as pointed out in Section 1, tidal wave is a mixed mode of surface gravity wave and inertial wave such that its frequency is beyond the range of $-2$ and 2. Secondly, negative tidal frequencies lead to stronger resonances than the corresponding positive frequencies (dashed lines are higher than solid lines of the same colour). This implies that rotation breaks the symmetry of tidal resonance.

Now, we study the effect of friction. Fig. 2(a) shows the tidal peak versus $\Omega$ at $\epsilon = 0.1$ with the four values of $\gamma/\sqrt{g/R} = 10^{-4}$, $10^{-3}$, $10^{-2}$ and $10^{-1}$. A comparison of different frictions indicates that stronger friction suppresses tidal resonances more, and at $\gamma/\sqrt{g/R} = 10^{-1}$ tidal resonances are heavily suppressed. Fig. 2(b) shows the tidal peak versus $\epsilon$ at $\omega = -1$ with the four values of $\gamma/\sqrt{g/R}$. In addition to showing similar results to Fig. 2(a), it provides the new information that relatively high tidal resonances with strong friction ($\gamma/\sqrt{g/R} = 10^{-2}$ and $10^{-1}$) appear at fast rotation $\epsilon > 0.1$ (i.e. the blue and green curves). Therefore, friction cannot completely suppress tidal resonances as long as rotation is sufficiently fast.

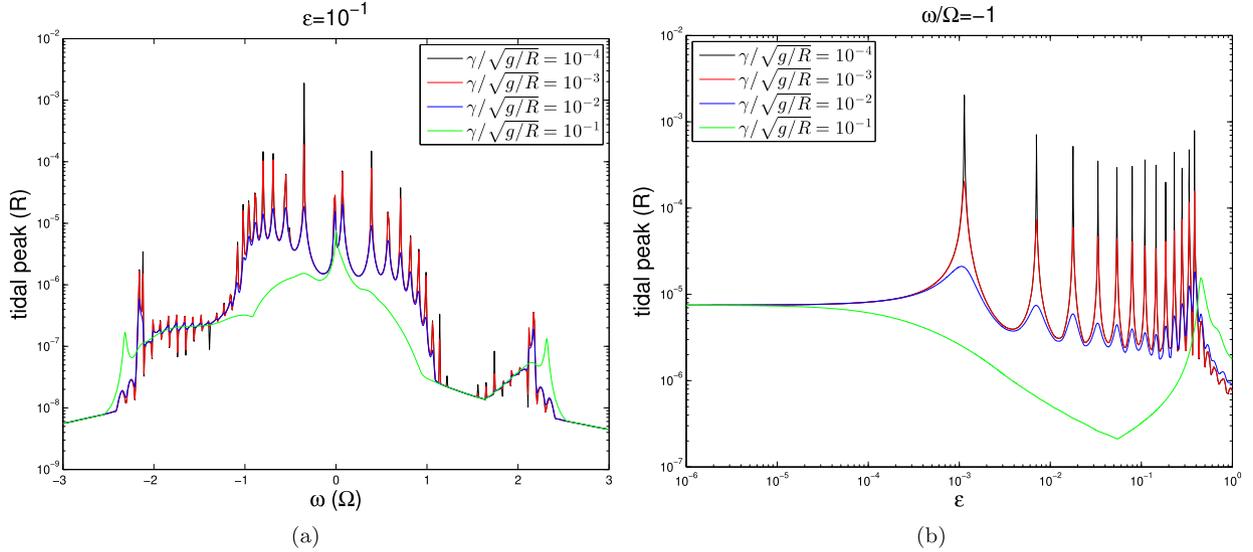

**Figure 2.** (a) Tidal peak versus $\omega$ with different $\gamma/\sqrt{g/R}$ at $\epsilon = 10^{-1}$. (b) Tidal peak versus $\epsilon$ with different $\gamma/\sqrt{g/R}$ at $\omega/\Omega = -1$.

To have a global view of tidal resonances, we plot the contours of the tidal peak as a function of $\omega$ and $\epsilon$ with the four values of $\gamma/\sqrt{g/R}$, as shown in Fig. 3. The red dots in curved stripes denote very strong tidal resonances. Similar to the above results, tidal resonances appear at fast rotation and concentrate within the frequency range $\omega = -1$ to 1, negative values of $\omega$ lead to more tidal resonances than positive values, and friction suppresses tidal resonances. Moreover, these contour figures provide the new information that a lower tidal frequency is required to generate the same level of tidal peak at a faster rotational rate (see the orientation of the curves).

To end this section, we study the effect of the thickness $h$ of the fluid layer on tidal resonances. We calculate $h = 2 \times 10^{-3}$ and $3 \times 10^{-3}$ (too large a value of $h$ breaks the shallow-water approximation) for comparison with $h = 1 \times 10^{-3}$. Fig. 4 shows the tidal peak versus $\omega$ (Fig. 4a) and versus $\epsilon$ (Fig. 4b) with different $h$ at $\gamma/\sqrt{g/R} = 10^{-3}$. The thickness of the fluid layer can change the tidal resonances, especially out of the range of the inertial wave (see the range $|\omega| > 2$ in Fig. 4a), because it changes the natural frequency of the tidal wave. However, this change becomes weaker when rotation becomes faster (see the differences in the three curves with increasing $\epsilon$ in Fig. 4b).

## 4 SUMMARY

In this paper, we study the tides on the surfaces of rapidly rotating stars or planets by numerically solving Laplace's tidal equations with friction terms. When $\epsilon$ reaches 0.1, a great amount of tidal resonances appear. Negative tidal frequency leads to stronger tidal resonances than the corresponding positive values. Friction suppresses tidal resonances but it cannot completely suppress them at fast rotation. To generate the same level of tidal peak, a faster rotational rate corresponds to a lower tidal frequency. The fluid layer thickness changes tidal resonances but this change is weaker when rotation is faster.

The non-linear effect is not considered in this paper. Kumar & Goodman (1996) pointed out that the non-linear effect damps waves through wave breaking and Wei (2016a) pointed out that the non-linear effect suppresses tidal resonances. Compared with the friction terms, it is not known whether the non-linear effect will be striking on the surface tide. Another effect is the magnetic field. Wei (2016b, 2018) and Lin & Ogilvie (2018) pointed out that the magnetic field can modify the dispersion relation of the tidal wave. These two effects on the surface tide of a rapidly rotating body will be explored in a future study.

## ACKNOWLEDGEMENTS

This work is financially supported by the 1000 Youth Talents programme of the Chinese government and the National Science Foundation of China (grant no. 11872246).

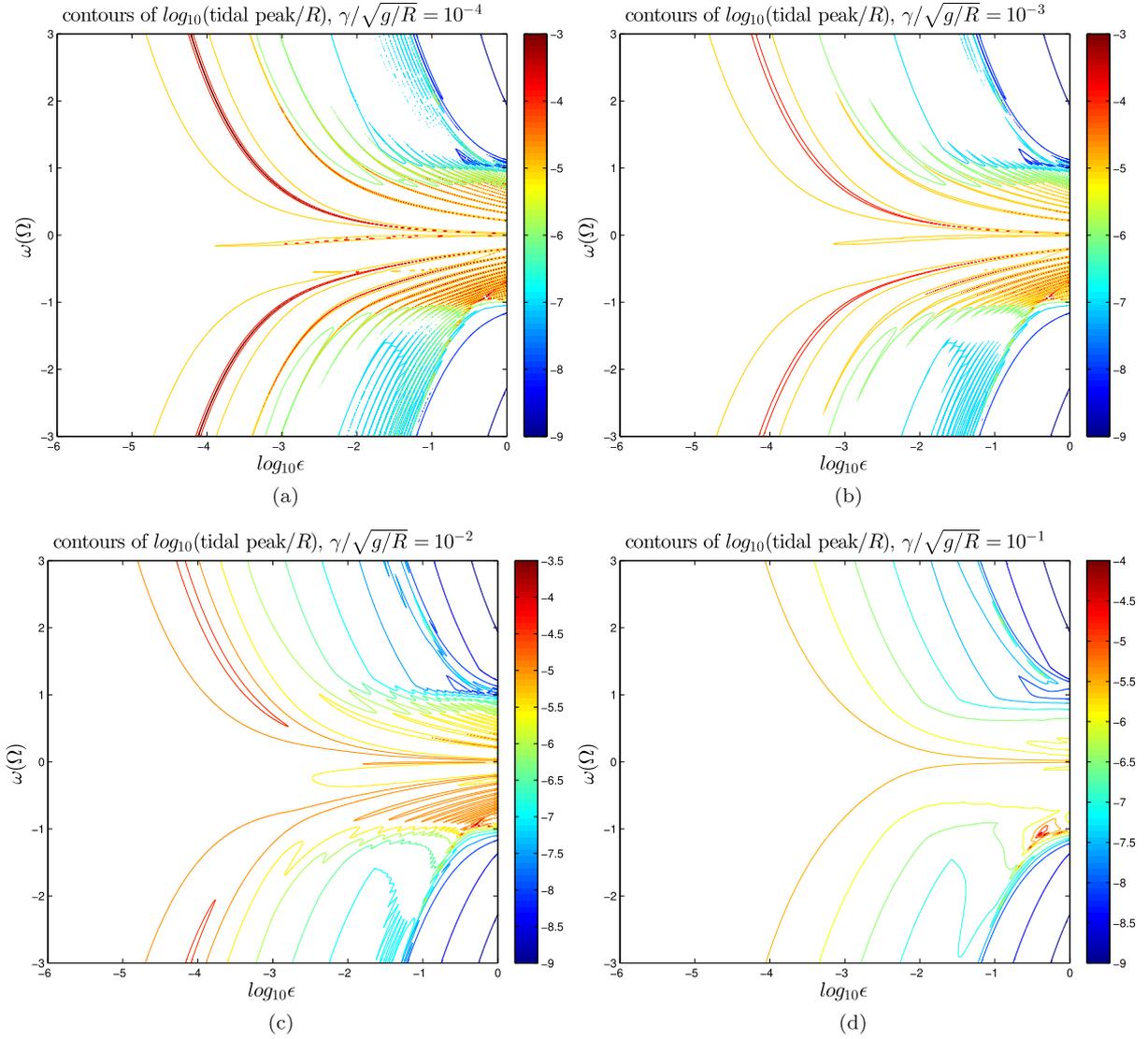

**Figure 3.** Contours of the tidal peak as a function of $\omega$ and $\epsilon$ with different $\gamma/\sqrt{g/R}$.

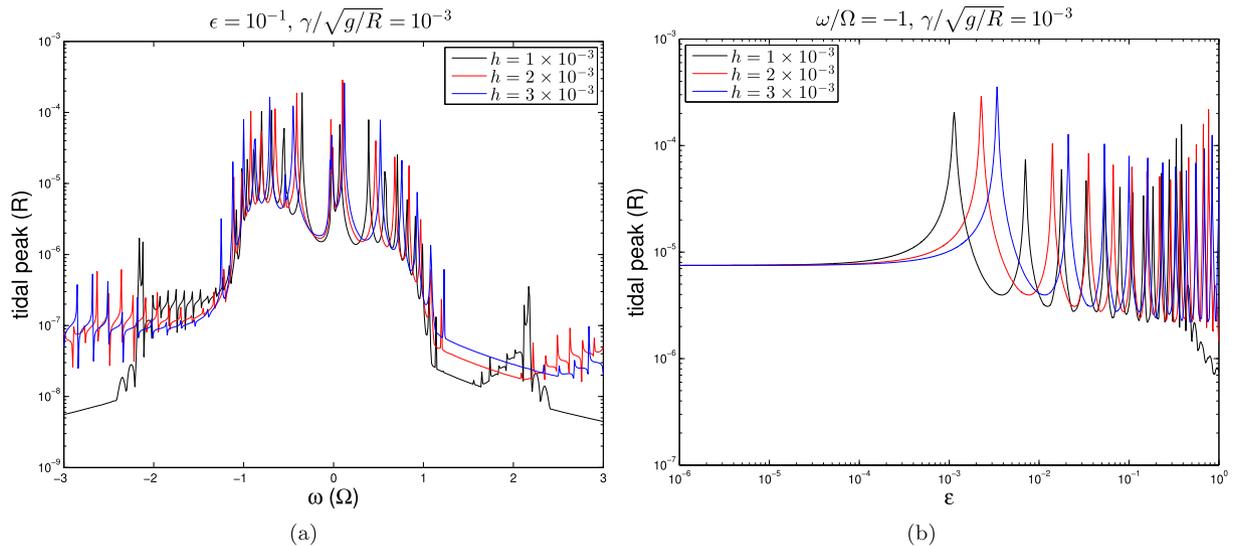

**Figure 4.** (a) Tidal peak versus $\omega$ with different values of $h$ at $\epsilon = 10^{-1}$ and $\gamma/\sqrt{g/R} = 10^{-3}$. (b) Tidal peak versus $\epsilon$ with different values of $h$ at $\omega/\Omega = -1$ and $\gamma/\sqrt{g/R} = 10^{-3}$.